\documentclass[aps,prb,twocolumn,superscriptaddress]{revtex4}
\usepackage{amsmath,epsfig,mathrsfs,graphicx,amssymb}

\def\be#1\ee{\begin{equation}#1\end{equation}}

\newcommand{\ba}{\begin{eqnarray} }
\newcommand{\ea}{\end{eqnarray} }

\begin{document}

\title{Proposal for a cumulant-based Bell test for mesoscopic junctions}

\author{Adam Bednorz}
\email{Adam.Bednorz@fuw.edu.pl}
%\altaffiliation[Electronic address: ]{Adam.Bednorz@fuw.edu.pl}
\affiliation{Fachbereich Physik, Universit{\"a}t Konstanz, D-78457 Konstanz, Germany}
\affiliation{Faculty of Physics, University of Warsaw, Ho\.za 69, PL-00681 Warsaw, Poland}
\author{Wolfgang Belzig}
\affiliation{Fachbereich Physik, Universit{\"a}t Konstanz, D-78457 Konstanz, Germany}
\date{\today}

\begin{abstract}

  The creation and detection of entanglement in solid state
  electronics is of fundamental importance for quantum information
  processing. We prove that second-order quantum correlations can be
  always interpreted classically and propose a general test of
  entanglement based on the 
  violation of a classically derived inequality for continuous
  variables by fourth-order quantum correlation functions.  Our
  scheme provides a way to prove the existence of entanglement in a
  mesoscopic transport setup by measuring higher order cumulants
  without requiring the additional assumption of a single charge
  detection.
\end{abstract}

\maketitle
\section{Introduction}

The quantum theory cannot be explained by any underlying classical local hidden
variable model, according to the Bell theorem.\cite{bel}  It allows to
verify that, contrary to
the classical case, the results of quantum mechanics violate a special inequality
\cite{bein} under the conditions: (i) dichotomy
of measurement outcomes or their restricted set in some generalizations,
\cite{cglmp} (ii)
freedom of choice of the measured observables \cite{will} and (iii) the time of
the choice and measurement of the observable shorter than the
communication time between the observers.  Relaxing any of the above
three conditions opens a (i) detection, (ii) free will or (iii)
communication loophole, which permits to construct a local hidden
variable model explaining the results of the experiments.\cite{loop}
The performed experimental tests confirmed the violation of Bell inequality with all the loopholes closed
\cite{belx} but never all simultaneously. \cite{loopfree} The
remaining loophole is "closed" by a reasonable additional assumption.
The Bell test is also stronger than the entanglement criterion, viz. the
nonseparability of states, \cite{ent} which assumes a finite dimension
of the Hilbert space.  Loophole-free violation of Bell inequality,
not just entanglement, is also necessary for successful quantum cryptography,\cite{gis06} although the loopholes are
probably less important than decoherence problems.
\cite{scalrev}

There is a growing interest in tests of nonclassicality in solid state
systems, especially electrons in noninteracting mesoscopic junctions.\cite{entsol,entrev,heiblum} 
Unlike bosons, even noninteracting
fermions in the Fermi sea can get entangled due to the Pauli exclusion
principle.  For example, entangled electron-hole pairs are created at
both sides of a biased tunnel junction.  So far the efforts
concentrated on testing entanglement by second-order current
correlations.\cite{excoop}  Unfortunately, to make the genuine Bell
test, the charge flow quantization must be measured directly, which has
so far only been achieved in quantum dots.\cite {dot}
However, in tunnel junctions and quantum point contacts rather current
cumulants \cite{fcs} are directly
accessible and, so far, the noise \cite{noi} and the third cumulant \cite{three}
of the current have been measured.  
The quantization of charge flow is also not so evident at short timescales or high
frequencies when vacuum fluctuations of the Fermi sea play a role.
\cite{vac}  In some cases, energy filters can restore the Bell
correlations at short times,\cite{han} at the expense of opening the
detection loophole (most of electrons get lost).\cite{fazio}

In this paper
we present a genuine Bell test for mesoscopic junctions, which does
not require the usual assumption that only quantized charge transfers
are detected.  Instead of quantized events, we shall treat the current
as a continuous, time-dependent observable. As we will show-second
order correlations in this case can be always explained
classically. Hence, we need a Bell inequality for unbounded variables,
without a sharp dichotomy, which will require to exploit correlation
functions and higher moments/cumulants. Such an inequality has
been recently discovered \cite{belcon} but the violation requires at
least 10 observers and 20th-order averages.  Moreover, the
corresponding Bell-like state is not feasible in mesoscopic junctions.
Our inequality will need only two observers and maximally fourth
moments/cumulants.  The inequality reduces to the usual Bell inequality
if the quantization is granted.  A violation is possible in a mesoscopic
tunnel junction with spin filtered leads or pierced by tunable
magnetic flux.

We first prove \emph{weak positivity} (classicality of second order quantum correlations),
next conctruct the Bell-type inequality based of fourth order moments, then implement it in the tunnel junction
and finally discuss possible loopholes.

\section{Weak positivity}

Let us begin with the simple proof that first- and second order
correlations functions can be always reproduced classically.  
To see this, consider a real symmetric correlation matrix
$2C_{ij}=2\langle A_iA_j\rangle=\mathrm{Tr}\hat{\rho}
\{\hat{A}_i,\hat{A}_j\}$ with $\{\hat{A},\hat{B}\}=\hat{A}\hat{B}+\hat{B}\hat{A}$
for arbitrary, even noncommuting
observables $\hat{A}_i$ and the density matrix $\hat{\rho}$. This includes
all possible first-order averages $\langle A_i\rangle$ by setting one
observable to identity or subtracting averages ($A_i\to A_i-\langle A_i\rangle$).
Since $\mathrm{Tr}\hat{\rho}\hat{X}^2\geq 0$ for
$\hat{X}=\sum_i\lambda_i\hat{A}_i$ with arbitrary real $\lambda_i$, we find that the correlation matrix $C$ is positive definite and any correlation can be simulated by a
classical Gaussian distribution $\rho\propto \exp(-\sum_{ij}{C^{-1}}_{ij}A_iA_j/2)$.
Note that the often used dichotomy $A=\pm 1$ is equivalent to $\langle
(A^2-1)^2\rangle=0$, which requires $\langle A^4\rangle$. Moreover, every classical inequality $\langle
(f(\{A_i\})^2\rangle\geq 0$ contains the highest correlator of even order.
 Hence, to detect nonclassical effects with unbounded observables, we have to consider the fourth moments.

\section{Bell-type inequality}

As usual we introduce two separate observers, Alice and Bob that are
free to choose between two observables, ($A$, $A'$) and ($B$, $B'$),
respectively.  The measurements can give arbitrary outcomes (not just
$\pm 1$).  We have the following algebraic identities:
\begin{eqnarray}
&&(AB+A'B+AB'-A'B')(A^2+A'^2+B^2+B'^2)\nonumber\\
&&
=2(A^3B+AB^3+A'^3B+A'B^3\nonumber\\
&&
+A^3B'+AB'^3-A'^3B'-A'B'^3)\label{iden2}\\
%&&+(AB+A'B+AB'-A'B')(A^2+B^2+A'^2+B'^2-4))\nonumber
 &&+AB[(A'^2-B^2)+(B'^2-A^2)]\nonumber\\
&&+A'B[(A^2-B^2)+(B'^2-A'^2)]\nonumber\\
&&+AB'[(A'^2-B'^2)+(B^2-A^2)]\nonumber\\
&&-A'B'[(A^2-B'^2)+(B^2-A'^2)]\nonumber
\end{eqnarray}
and
\begin{eqnarray}
&&(AB+A'B+AB'-A'B')^2=(A^2+A'^2)(B^2+B'^2)\nonumber\\
&&
+AA'[(B^2-A^2)+(B^2-A'^2)]\nonumber\\
&& -AA'[(B'^2-A^2)+(B'^2-A'^2)]\nonumber\\
&&+BB'[(A^2-B^2)+(A^2-B'^2)]\nonumber\\
&&-BB'[(A'^2-B^2)+(A'^2-B'^2)].\label{iden} 
\end{eqnarray}
We now apply the Cauchy inequality $2|\langle xy\rangle| \leq \langle x^2+y^2\rangle$ to $x=AB+A'B+AB'-A'B'$, $y=(A^2+A'^2+B^2+B'^2)/2$,
then $|\langle X\rangle|\leq |\langle X+Y\rangle|+|\langle Y\rangle|$
for $X+Y$ given by the right-hand side of (\ref{iden2}) and
$X=2(A^3B+AB^3+A'^3B+A'B^3
+A^3B'+AB'^3-A'^3B'-A'B'^3)$. Then we use again the Cauchy inequality for $x,y=A^2,A'^2,B^2,B'^2$.
On the right-hand sides of (\ref{iden2}) and (\ref{iden}) we have 16
terms of the form $\pm\langle AA'(B^2-A'^2)\rangle$. To decouple those
containing simultaneous measurements of $A(B)$ and $A'(B')$ we first
use the triangle inequality 
$|\langle\sum_i x_i\rangle|\leq \sum_i|\langle x_i\rangle|$ for the
sum of all
terms and finally the Cauchy-Bunyakovsky-Schwarz-type inequality
$|\langle AA'(A^2-B^2)\rangle|\leq \langle A^4\rangle^{1/4}\langle A'^4\rangle^{1/4}\langle(A^2-B^2)^2\rangle^{1/2}$
to each term individually. 
We end up with our main inequality 
\begin{eqnarray}
&&|\langle AB(A^2+B^2)\rangle+
  \langle A'B(A'^2+B^2)\rangle\nonumber\\
&&+\langle AB'(A^2+B'^2)\rangle-\langle A'B'(A'^2+B'^2)\rangle|/2
\leq
\label{bell} \\
&& \left(\langle A^4\rangle+\langle A'^4\rangle+\langle
  B^4\rangle+\langle B'^4\rangle\right)/2\nonumber + \\ && \frac{1}{4}
\sum_{C,D,E=\{A,A',B,B'\}}^{D\neq C;E\neq C,D,D'}
\sqrt{\sqrt{\left\langle C^4\right\rangle}\sqrt{\langle D^4\rangle}\left\langle\left(D^2-E^2\right)^2\right\rangle},\nonumber
\end{eqnarray}
where $D'=A(B)$ when $D=A'(B')$. For the complete derivation, see the Appendix. 

The inequality contains up to fourth-order averages
which is a trade-off for relaxing the condition of dichotomy (or trichotomy, considering also $0$).  It reduces to the standard
Bell inequality
\begin{equation}
|\langle AB\rangle+\langle A'B\rangle+\langle AB'\rangle-\langle A'B'\rangle|\leq 2
\end{equation}
if we restrict the possible values of
$A$,$A'$,$B$,$B'$ to $\pm 1$.  
If all observables are allowed to take the additional value $0$ only
simultaneously then the inequality still reproduces Bell multiplied by
the probability of nonzero outcomes.  All correlations in the
inequality are measurable also in a Bell-type test, because none of
them contains $AA'$ or $BB'$.  Hence, we can say that the degrees of
freedom measured by $A^{(\prime)}$ and $B^{(\prime)}$ are entangled if
the inequality (\ref{bell}) of their correlators is violated. We
emphasize that this requires only the assumption of nonnegative probability distribution $\rho(A,A',B,B')\geq 0$ and provides an
unambiguous proof for entanglement.

Returning to quantum mechanics, let us take the standard Bell state \cite{bein}
$\hat{\rho}=(\hat{1}-\hat{\boldsymbol\sigma}_A\cdot\hat{\boldsymbol\sigma}_B)/4$, 
$\boldsymbol\sigma=(\sigma_1,\sigma_2,\sigma_3)$, with $\hat{\sigma}_i$ --
standard spin Pauli matrices $\{\hat{\sigma}_i,\hat{\sigma}_j\}=2\delta_{ij}\hat{1}$ acting in Hilbert space $\mathcal H_A\otimes\mathcal H_B$, $\hat{A}^{(\prime)}=\boldsymbol a^{(\prime)}\cdot\hat{\boldsymbol{\sigma}}_A$, $|\boldsymbol a^{(\prime)}|=1$
($A\leftrightarrow B$) and averages $\langle A^{(\prime)n}B^{(\prime)m}\rangle=\mathrm{Tr}\hat{\rho}\hat{A}^{(\prime)n}\hat{B}^{(\prime)m}$. 
In particular $\langle A^{(\prime)4}\rangle=\langle B^{(\prime)4}\rangle=\langle A^{(\prime)2}B^{(\prime)2}\rangle=1$ and $\langle A^{(\prime)}B^{(\prime)3}\rangle
=\langle A^{(\prime)3}B^{(\prime)}\rangle=-\boldsymbol a^{(\prime)}\cdot\boldsymbol b^{(\prime)}$. The inequality (\ref{bell}) is violated as it reads $2\sqrt{2}\leq 2$
for $\boldsymbol a'$,$\boldsymbol b$,$\boldsymbol
a$,$\boldsymbol b'$ in one plane at angles $0$, $\pi/4$,
$\pi/2$, $3\pi/4$, respectively.

\section{Test on tunnel junction}

Now we implement the Bell example in a beam splitting device involving
fermions scattered at a tunnel junction.  The junction is described by fermionic operators
around the Fermi level.\cite{blanter}  Each operator
$\hat{\psi}_{A\bar n}$ is denoted by the mode number $n\in\{1..N\}$
and the spin orientation $\sigma$, $\bar n=(n,\sigma)$ and $A=L,R$ for left
and right going electrons, respectively.  Each mode has its own Fermi
velocity $v_n$ and transmission coefficient $\mathcal T_n$ (reflection
$\mathcal R_n=1-\mathcal T_n$).  We will assume noninteracting
electrons and energy- and spin-independent transmission through the
junction.  The Hamiltonian is
\begin{eqnarray}
  &&\hat{H}=\sum_{\bar n}\int dx\left\{i\hbar v_n[\hat{\psi}^\dag_{L\bar
    n}(x) \partial_x\hat{\psi}_{L\bar n}(x)- L\leftrightarrow R]\right.\nonumber\\
&&+eV\theta(-x)[\hat{\psi}^\dag_{L\bar n}(x)\hat{\psi}_{L\bar n}(x)+L\leftrightarrow R]\nonumber\\
  &&\left.+q_n\delta(x)[\hat{\psi}^\dag_{L\bar n}(x)\hat{\psi}_{R\bar n}(-x)+
    \hat{\psi}^\dag_{R\bar n}(x)\hat{\psi}_{L\bar n}(-x)]
  \right\}\,.\label{hhh}
\end{eqnarray}
The fermionic operators satisfy anticommutation relations
$\{\hat{\psi}_{a}(x),\hat{\psi}_{b}(x')\}=0$ and
$\{\hat{\psi}_{a}(x),\hat{\psi}^\dag_{b}(x')\}=\delta_{ab}\delta(x-x')$
for $a,b=L\bar n,R\bar m$. The transmission coefficients are $\mathcal
T_n=\cos^2(q_n/\hbar v_n)$. The system's current operator is defined
$\hat{I}_n(x)=\sum_{\sigma}ev_n\hat{\psi}^\dag_{L\bar n}(x)
\hat{\psi}_{L\bar n}(x)-L\leftrightarrow R$ and the density matrix is
$\hat{\rho}\propto\exp(-\hat{H}/k_BT)$.
 
\begin{figure}
\includegraphics[scale=.4,angle=0]{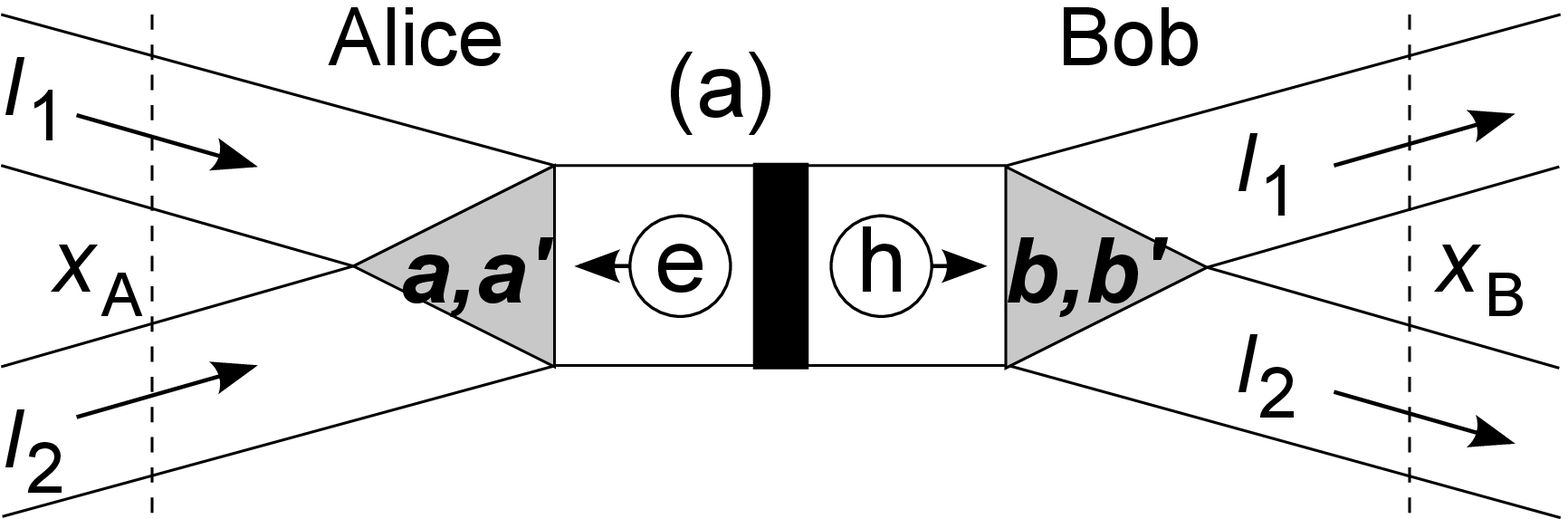}\\
\medskip
\includegraphics[scale=.4,angle=0]{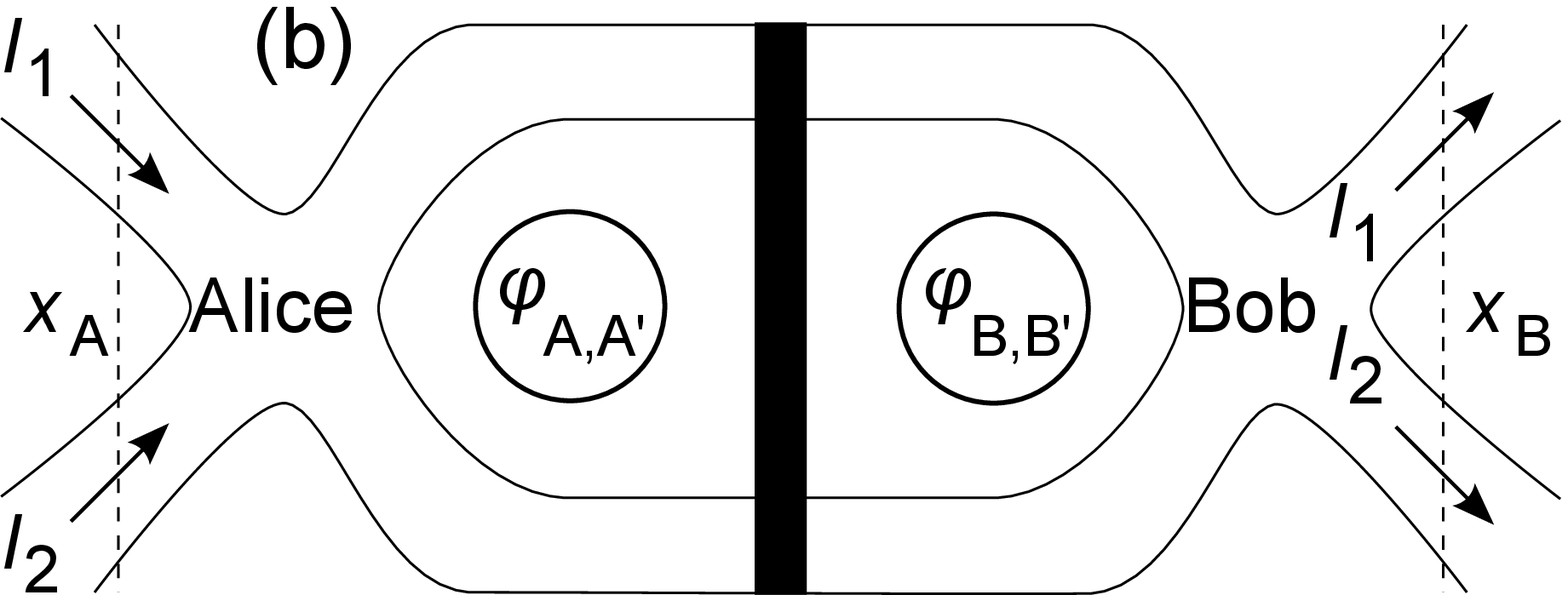}
\caption{Proposals of experimental setups for the Bell test. In both cases the black bar represents the scattering barrier, producing entangled electron-hole pairs. The tested observable is the difference
of currents, $I_1-I_2$, at left (Alice) or right (Bob) side. The correlations 
depend on the spin scattering (a) or magnetic fluxes (b).}\label{setu}
\end{figure}

The Bell measurement will be performed by adding spin filters or
magnetic flux at both sides of the junctions as shown in Fig. \ref{setu}.
In both cases we have to add $\hat{H}'=\sum_{ab}
\int dx\; eV_{ab}(x)\hat{\psi}^\dag_{a}(x)\hat{\psi}_{b}(x)$
to the Hamiltonian (\ref{hhh}) where $V_{ab}(x)$ is scattering potential,
localized near detectors.
The effect of each part of the Hamiltonian on single-mode wave function can be described by
three scattering matrices \cite{blanter}
\be
s_i=\left(\begin{array}{cc}
r_i&t_i\\
t'_i&r'_i\end{array}\right),
\ee
where $i=A,T,B$ describe scattering at the left detector,
junction and the right detector, respectively.  The junction
has diagonal transmission and reflection submatrices with $t_T=t'_T=i\sqrt{\mathcal
  T}\hat{1}$. 
In the case of spin filters we assume 
$4\times 2$ transmission
$
2t_A=(1+\boldsymbol a\cdot\hat{\boldsymbol\sigma}\:\: 1-\boldsymbol a\cdot\hat{\boldsymbol\sigma}),
$
where $|\boldsymbol a|=1$. Alternatively, having a tunable geometry of the
scatterer, we could introduce an
"artificial spin" filter taking $\sigma_{1,3}$ acting in the mode
space instead of spin space. For magnetic fluxes $r_A=0$ and
\be
t_A=\left(\begin{array}{cc}
e^{i\phi_A}&0\\
0&1\end{array}\right)\frac{1}{\sqrt{2}}\left(\begin{array}{cc}
1&1\\
1&-1\end{array}\right).
\ee
where $\phi_A$ represents Aharonov-Bohm phase picked on the upper branch.
The matrices can be enlarged to represent the $2N$-mode junction.
In both cases, the transmission coefficients for the total scattering matrix are $\mathcal
T_{11}=\mathcal T_{22} =\mathcal T(1+\boldsymbol a\cdot\boldsymbol
b)/2$ and $\mathcal T_{12}=\mathcal T_{21} =\mathcal T(1-\boldsymbol
a\cdot\boldsymbol b)/2$, where $\boldsymbol a=(\cos\phi_A,\sin\phi_A,0)$
in the case of magnetic fluxes.

As in the previous proposals \cite{entsol,entrev} the tunnel barrier produces electron-hole pairs with
entangled spins or orbitals. Alice and Bob can test the inequality (\ref{bell}) by measuring the difference between charge fluxes in the upper and lower arm as shown in Fig. \ref{setu}. For Alice, the measured observable reads in the Heisenberg picture
\be
\hat{A}=\int dt\:f(t)(\hat{I}_1(x_A,t)-\hat{I}_2(x_A,t))/e.
\ee
for the filter setting $\boldsymbol a$. 
Here $x_A$ is the point of measurement, satisfying
$\mathrm{max}\{|eV|,k_BT\}|x_A/v_n\hbar|\ll \mathcal T$ with $f(t)$ slowly changing on the timescale
$\hbar/\mathrm{max}\{|eV|,k_BT\}$. One defines analogically $A'$ for $\boldsymbol a'$ and
$B$, $B'$ for Bob.  

The measured probability distribution can be treated as a convolution
$\rho(A,B)=\int dA'dB'\rho_d(A-A',B-B')\varrho(A',B')$, where $\rho_d$ is the Gaussian detection
noise (independent of the system and later subtracted)
and $\varrho$ is a quasiprobability,\cite{our} where
\begin{eqnarray}
&&\langle X_1(t_1)\cdots X_n(t_n)\rangle_\varrho=\\
&&\mathrm{Tr}\hat{\rho}\{\hat{X}_1(t_1),\{\cdots\{\hat{X}_{n-1}(t_{n-1}),
\hat{X}_n(t_n)\}\cdots\}\}/2^{n-1}\nonumber
\end{eqnarray}
for time ordered observables, $t_1\leq t_2\leq\cdots\leq t_n$.
The detection noise adds $I_{di}$ to the measurement outcome $I_i$ with $\langle
I_{di}(0) I_{dj}(t)\rangle=e^2\delta(t)\delta_{ij}/\tau$. In the
interaction-free limit [the sensitivity
$\tau$ much smaller than the time resolution of the
measurement, the timescale of $f(t)$] one can calculate averages with respect to $\varrho$ using
existing methods based on full counting statistics and its extension.
\cite{entsol,entrev,fcs,blanter, zaikin1}
The averages needed in the inequality (\ref{bell}) can be derived using four-lead full counting statistics generating functional\cite{fcs}
in the tunneling limit ($\mathcal T\ll 1$),
\begin{eqnarray}
&&S(\chi_{A},\chi_B)=\ln\langle e^{iA\chi_A+iB\chi_B}\rangle_\varrho\nonumber\\
&&
=N\mathcal T\hbar^{-1}\sum_{\alpha,\beta=\pm 1}\sum_{\pm}\int dt f(t)\int dE \times\label{fcss}\\
&&f_\pm(E)(1-f_\mp(E))(1+\alpha\beta\boldsymbol a\cdot\boldsymbol b)(e^{\pm i\alpha\chi_{A}\pm i\beta\chi_{B}}-1)\nonumber
\end{eqnarray}
with Fermi distributions $f_{\pm}(E)=(1+e^{(E\pm eV/2)/k_BT})^{-1}$. In our case, one obtains a
simple physical picture: the electron-hole Bell pairs are transmitted according to
Poissonian statistics. The averages (cumulants and moments) are found by taking derivatives of (\ref{fcss}) with respect to $\chi$.
In particular,
we have $
\langle A^{(\prime)}\rangle_\varrho=
\langle B^{(\prime)}\rangle_\varrho=0$ and $\langle A^{(\prime)}B^{(\prime)3}\rangle_\varrho=\langle A^{(\prime)3}B^{(\prime)}\rangle_\varrho=({\boldsymbol a}^{(\prime)}\cdot{\boldsymbol
  b}^{(\prime)}) \langle A^4\rangle_\varrho$, $ \langle A^4\rangle_\varrho=\langle A^{(\prime) 4}\rangle_\varrho=\langle B^{(\prime)4}\rangle_\varrho$
and $\langle A^2B^2\rangle_\varrho=\langle A^4\rangle_\varrho-2(1-(\boldsymbol a\cdot\boldsymbol b)^2))\langle A^2\rangle^2_\varrho$.  The inequality (\ref{bell})
gets a simplified form in this particular case,
\begin{equation}
|C(\boldsymbol a,\boldsymbol b,\boldsymbol a',\boldsymbol b')|
\leq 2+2\sum_{\boldsymbol d=\boldsymbol a,\boldsymbol a'}^{\boldsymbol e=\boldsymbol b,\boldsymbol b'}
\sqrt{(1- (\boldsymbol d\cdot \boldsymbol e)^2)\langle A^2\rangle^2_\varrho/\langle A^4\rangle_\varrho},\label{ineq}
\end{equation}
where $C(\boldsymbol a,\boldsymbol b,\boldsymbol a',\boldsymbol b')=\boldsymbol a\cdot\boldsymbol b+\boldsymbol a'\cdot\boldsymbol b+
\boldsymbol a\cdot\boldsymbol b'-\boldsymbol a'\cdot\boldsymbol b'$.
We stress that (\ref{ineq}) follows from theoretical predictions and
the experimental test still requires 
the measurement of all averages in (\ref{bell}).
We choose  $f(t)=\theta_\delta(2t_0-|t|)$, where $\theta_\delta(t)=\theta(t)$
for $|t|\gg\delta$ with a smooth crossover at $|t|\lesssim \delta$.

Having assumed the tunneling limit ($\mathcal T\ll 1$), we make the following approximations
\be
1/N\mathcal T\gg t_0\mathrm{max}\{|eV|,k_BT\}/h\gg t_0/\delta\gg 1\label{scales}
\ee
with $2N$ denoting the total number modes going through the 
barrier.  In this limit, all moments and cumulants are equal
\be
\langle A^{(\prime)2n}\rangle_\varrho\simeq \frac{2eVN\mathcal T t_0}{h}\coth\left(\frac{eV}{2k_BT}\right),\: n>0\label{avvg}
\ee
Hence the last term on the right-hand side of Eq. ~(\ref{ineq}) is
negligible and the inequality takes the usual form $|C(\boldsymbol a,\boldsymbol b,\boldsymbol a',\boldsymbol b')|
\leq 2$, which can be violated by appropriate choice of the spin axes.
 Instead of time domain,
one can measure correlations in frequency domain (up to $\omega\sim
1/t_0$) and make a Fourier transform.\cite{han} If the scattering is
mode-independent then one can assume that the junction consists of
minimally $N_0\simeq Gh/e^2$ independent channels, where $G=2N\mathcal Te^2/h$ is the
total conductance of the junction, and repeat the whole above
reasoning with $N$ replaced by $N/N_0$ (experimentally -- dividing
measured cumulants $\langle\langle\cdots\rangle\rangle$ by $N_0$).

\section{Loopholes}

The communication loophole is
still open, not only because the system is nonrelativistic but also
because the measurement time $t_0$ is larger than the flight time
between detectors ($|x_A-x_B|/v_n$). Let us 
impose shorter measurement $t_0\ll |x_A-x_B|/v_n$. Far from the barrier the vacuum
fluctuations of incoming and reflected current
do not cancel each other.  For the noise
measured at frequencies $\omega\gg v_n/|x_A|$, the incoming and
reflected current become independent and the noise saturates to the
same value as for the completely open barrier,
\begin{equation}
\int dt\:e^{i\omega t}\mathrm{Tr}\{\hat{I}_i(t),\hat{I}_j(0)\}\hat{\rho}/2
=\frac{N\delta_{ij}\omega}{2\pi}\coth\left(\frac{\hbar\omega}{2k_BT}\right).
\end{equation}
Hence for $t_0\ll
|x_A|/v_n$, we have
\be
\langle A^2\rangle_\varrho\simeq 2N\pi^{-2}\ln\left[\frac{\sinh(\pi k_BTt_0/\hbar)}{
\sinh(\pi k_BT\delta/\hbar)}\right],
\ee 
and $\langle A^4\rangle\simeq 3\langle A^2\rangle^2$,
which ruins any 
attempt to violate (\ref{bell}), as the tunneling factor $\mathcal T$ is lost.

Finally, the detection loophole is closed only partially, because we
get the violation of (\ref{bell}) only for the quasiprobability $\varrho$, after
subtraction of the detection noise $2t_0/\tau\gg 1$, which adds up to
$\langle A^2\rangle_\varrho$, and appears in the cumulant generating function 
$\ln\langle e^{i\chi_AA+i\chi_BB}\rangle_\rho
=\ln\langle e^{i\chi_AA+i\chi_BB}\rangle_\varrho -(\chi^2_A+\chi_B^2)t_0/\tau$, preventing (\ref{bell}) from violations.
Such subtraction is justified because the Gaussian detection noise is independent of the system
as an inherent feature of interaction-free measurement
and can be experimentally measured at zero bias voltage.\cite{vac}
It affects only the second cumulant not higher cumulants, for example
$\langle \delta X\delta Y\delta Z\delta W\rangle
-\langle \delta X\delta Y\rangle\langle \delta Z\delta W\rangle-
\langle \delta X\delta Z\rangle\langle \delta Y\delta W\rangle-
\langle \delta X\delta W\rangle\langle \delta Z\delta Y\rangle$ is the same
for $\varrho$ and $\rho$ with $X,Y,Z,W=A,B$,
which can be confirmed experimentally.

\section{Conclusions}

We have proved that second-order quantum correlations can be always interpreted classically.
We constructed a classical inequality for nonlocal correlation
measurements involving up to fourth-order correlations.
A violation of this inequality can serve as a cumulant-based Bell test for entanglement.
In particular, it can be applied to  mesoscopic
junctions relaxing the charge quantization assumption.
 A spin-resolved
quantum measurement on tunnel junctions violates the inequality in an
experimentally accessible range of temperatures, voltages and
time/frequency resolution.  The communication and detection loophole
remain open due to long-distance vacuum fluctuations and detection
noise. Closing these loopholes will be a challenge for future research. Nevertheless, the experimental
violation of the inequality (\ref{bell}) will be a very important step
for the understanding and control of quantum entanglement in mesoscopic
physics.

\section*{Acknowledgments} 
We are grateful for fruitful discussions with J. Gabelli, B. Reulet, and R. Fazio.
Financial support from the DFG through SFB 767 and SP 1285 is acknowledged.

\section*{Appendix}
\renewcommand{\theequation}{A\arabic{equation}}
\setcounter{equation}{0}

We summarize some details on the derivation of our main inequality (\ref{bell}) ,
following the instructions in the main article.
Let us first rewrite identity (\ref{iden2}) as
\begin{widetext}
\begin{eqnarray}
&&\underbrace{(AB+A'B+AB'-A'B')(A^2+A'^2+B^2+B'^2)}_{Z=X+Y}\nonumber\\
&&=\underbrace{2(A^3B+AB^3+A'^3B+A'B^3+A^3B'+AB'^3-A'^3B'-A'B'^3)}_X\label{eq:iden1}\\
 &&+\underbrace{AB[(A'^2-B^2)+(B'^2-A^2)]+A'B[(A^2-B^2)+(B'^2-A'^2)]}_{Y}\nonumber\\
&&+\underbrace{AB'[(A'^2-B'^2)+(B^2-A^2)]-A'B'[(A^2-B'^2)+(B^2-A'^2)]}_{\textrm{still
    } Y},\nonumber
\end{eqnarray}
where we labeled the terms for later use.
Now we start with the derivation of the main inequality. We use the
basic inequality 
\begin{equation}
  \label{eq:main}
  |\langle X\rangle|\leq |\langle X+Y\rangle|+|\langle Y\rangle|,
\end{equation}
with $X$ and $Y$ defined in Eq. (\ref{eq:iden1}). Note that $|\langle
X\rangle|$ is already the left hand side of the final inequality and
can be written in the more transparent expression $|\langle
X\rangle|$=$2|\langle AB(A^2+B^2)\rangle+ \langle
A'B(A'^2+B^2)\rangle+\langle AB'(A^2+B'^2)\rangle-\langle
A'B'(A'^2+B'^2)\rangle|$.
We next apply the Cauchy inequality from the paper to $Z=X+Y=xy$ with
$x=AB+A'B+AB'-A'B'$ and $y=(A^2+A'^2+B^2+B'^2)/2$, which gives
\begin{equation}
  \label{eq:1}
  |\langle(AB+A'B+AB'-A'B')(A^2+B^2+A'^2+B'^2)\rangle|
  \leq
  \langle (AB+A'B+AB'-A'B')^2\rangle+\frac 14 \langle(A^2+B^2+A'^2+B'^2)^2\rangle.
\end{equation}
Using the identity (\ref{iden}) for the first term of the right-hand side of (\ref{eq:1}) we find
\begin{eqnarray}
&&\langle(AB+A'B+AB'-A'B')^2\rangle=\langle(A^2+A'^2)(B^2+B'^2)\rangle
\label{eq:4}\\
&&
+\langle AA'(B^2-A^2)\rangle+\langle AA'(B^2-A'^2)\rangle 
-\langle AA'(B'^2-A^2)\rangle-\langle AA'(B'^2-A'^2)\rangle\nonumber\\
&&
+\langle BB'(A^2-B^2)\rangle+\langle BB'(A^2-B'^2)\rangle
-\langle BB'(A'^2-B^2)\rangle-\langle BB'(A'^2-B^2)\rangle.\nonumber
\end{eqnarray}
Now we collect all terms which consist of terms of the form $C^2D^2$ on
the right-hand side of (\ref{eq:1}) taking into account (\ref{eq:4}) and apply
Cauchy inequality to them
\begin{eqnarray}
&&\langle(A^2+A'^2)(B^2+B'^2)\rangle+\langle(A^2+B^2+A'^2+B'^2)^2\rangle/4
=\langle A^2B^2\rangle+\langle A'^2B^2\rangle+\langle
A^2B'^2\rangle+\langle A'^2B'^2\rangle
\nonumber\\
&&
+\left(\langle A^4\rangle+\langle B^4\rangle+\langle A'^4\rangle+\langle B'^4\rangle\right)/4+\left(\langle A^2B^2\rangle+\langle A^2A'^2\rangle+\langle A^2B'^2\rangle+\langle B^2A'^2\rangle+
\langle B^2B'^2\rangle+\langle A'^2B'^2\rangle\right)/2
\nonumber\\
&&\leq \left(\langle A^4\rangle+\langle B^4\rangle+\langle A^4\rangle+\langle B'^4\rangle+
\langle A'^4\rangle+\langle B^4\rangle+\langle A'^4\rangle+\langle
B^4\rangle\right)/2+\left(\langle A^4\rangle+\langle B^4\rangle+\langle
A'^4\rangle+\langle B'^4\rangle\right)/4
\nonumber\\\
&&+\left(\langle A^4\rangle+\langle B^4\rangle+\langle A^4\rangle+\langle A'^4\rangle+\langle A^4\rangle
+\langle B'^4\rangle+\langle B^4\rangle+\langle A'^4\rangle+
\langle B^4\rangle+\langle B'^4\rangle+\langle A'^4\rangle+\langle
B'^4\rangle\right)/4
\nonumber\\\label{eq:6}
&&=2\left(\langle A^4\rangle+\langle B^4\rangle+\langle A'^4\rangle+\langle B'^4\rangle\right).
\end{eqnarray}
Now we apply the triangular inequality to $|\langle Y\rangle|$ and get
\begin{eqnarray}
\label{eq:3}
&&|\langle AB(A'^2-B^2)\rangle+\langle AB(B'^2-A^2)\rangle+
\langle A'B(A^2-B^2)\rangle+\langle A'B(B'^2-A'^2)\rangle\nonumber\\
&&+\langle AB'(A'^2-B'^2)\rangle+\langle AB'(B^2-A^2)\rangle-
\langle A'B'(A^2-B'^2)\rangle-\langle A'B'(B^2-A'^2)\rangle|\nonumber\\
&&\leq|\langle AB(A'^2-B^2)\rangle|+|\langle AB(B'^2-A^2)\rangle|+
|\langle A'B(A^2-B^2)\rangle|+|\langle A'B(B'^2-A'^2)\rangle|\nonumber\\
&&+|\langle AB'(A'^2-B'^2)\rangle|+|\langle AB'(B^2-A^2)\rangle|+
|\langle A'B'(A^2-B'^2)\rangle|+|\langle A'B'(B^2-A'^2)\rangle|.
\end{eqnarray}
Finally, we apply the Cauchy-Bunyakovsky-Schwarz inequality to all relevant terms in (\ref{eq:3}) and (\ref{eq:4}) in
the form
\begin{equation}
|\langle CD(D^2-E^2)\rangle|\leq[\langle C^2D^2\rangle\langle(D^2-E^2)^2\rangle]^{1/2}\leq
\langle C^4\rangle^{1/4}\langle D^4\rangle^{1/4}\langle(D^2-E^2)^2\rangle^{1/2}
\end{equation}
and obtain 
\begin{eqnarray}
&&2|\langle AB(A^2+B^2)\rangle+
  \langle A'B(A'^2+B^2)\rangle+\langle AB'(A^2+B'^2)\rangle-\langle A'B'(A'^2+B'^2)\rangle|
\leq
\label{bell1} \\
&& 2\sum_{C}\langle C^4\rangle
 +\sum_C\sum_{D\neq C;E\neq C,D,D'}\{CDE\},\nonumber
\end{eqnarray}
\end{widetext}
equivalent to (\ref{bell}),
where $\{CDE\}=\langle C^4\rangle^{1/4}\langle
D^4\rangle^{1/4}\langle(D^2-E^2)^2\rangle^{1/2}$. The second sum in the last
term of the main inequality is understood as
\begin{equation}
  \label{eq:5}
  \sum_{C=\{A,A',B,B'\}}
  \sum_{D=\{A,A',B,B'\}}^{D\neq C}
  \sum_{E=\{A,A',B,B'\}}^{E\neq C,D,D'}\,,
\end{equation}
where $D''=D$. This term has therefore 16 summands, for example 4 of
the type $\{A,A',B\},\{A,A',B'\},\{A,B,A'\},\{A,B',A'\}$ for $C=A$ and
correspondingly for the other values of C.

\end{document}